\documentclass[twoside,journal]{IEEEtran}
\ifCLASSINFOpdf
\else
\fi
\usepackage[keeplastbox]{flushend}
\usepackage{cite}
\usepackage{hyperref}
\usepackage{subcaption} 
\usepackage{graphicx}
\usepackage{caption}
\usepackage[table,xcdraw]{xcolor}
\usepackage{booktabs,graphicx}
\usepackage{multirow}

\begin{document}
%
\title{Automatic Arm Motion Recognition Based on Radar Micro-Doppler Signature Envelopes}
%
%
%

\author{Zhengxin~Zeng,
        Moeness~Amin,~\IEEEmembership{Fellow,~IEEE,}
        and~Tao~Shan,~\IEEEmembership{Member,~IEEE}
\thanks{The work of Mr. Zhengxin is funded by the International Graduate Exchange Program of Beijing Institute of Technology, and was performed while he was a Visiting Scholar at the Center for Advanced Communications, Villanova University.}
\thanks{An earlier version of this paper was presented at the 2019 IEEE Radar Conference and was published in its Proceedings: \url{https://ieeexplore.ieee.org/stamp/stamp.jsp?tp=&arnumber=8835661}}
\thanks{Z. Zeng and T. Shan are with the School of Information and Electronics, Beijing Institute of Technology, Beijing, China (e-mail: 3120140293@bit.edu.cn; shantao@bit.edu.cn).}
\thanks{M. Amin is with Center for Advanced Communications, Villanova University, Villanova, PA 19085, USA (e-mail: moeness.amin@villanova.edu).}}

%
%

\markboth{IEEE SENSORS JOURNAL, VOL. XX, NO. XX MONTH X, XXXX}
{Zeng \MakeLowercase{\textit{et al.}}: Automatic Arm Motion Recognition Based on Radar Micro-Doppler Signature Envelopes}
%



\maketitle

\begin{abstract}
In considering human-machine interface (HMI) for smart environment, a simple but effective method is proposed for automatic arm motion recognition with a Doppler radar sensor. Arms, in lieu of hands, have stronger radar cross-section and can be recognized from relatively longer distances. An energy-based thresholding algorithm is applied to the spectrograms to extract the micro-Doppler (MD) signature envelopes. The positive and negative frequency envelopes are concatenated to form a feature vector. The nearest neighbor (NN) classifier with Manhattan distance (L1) is then used to recognize the arm motions. It is shown that this simple method yields classification accuracy above 97 percent for six classes of arm motions. Despite its simplicity, the proposed method is superior to those of handcrafted feature-based classifications and low-dimension representation techniques based on principal component analysis (PCA), and is comparable to convolutional neural network (CNN).
\end{abstract}
\begin{IEEEkeywords}
Arm motion recognition, Doppler radar, micro-Doppler signature, spectrograms.
\end{IEEEkeywords}

\IEEEpeerreviewmaketitle

\section{Introduction}
\IEEEPARstart{R}{ardar} has become of increased interest for indoor sensing monitoring, including home security, smart homes, assisted living, elderly care, and medical diagnosis applications \cite{amin2017radar,amin2016radar,seifert2019detection}. Over the past decade, much work has been done in human motion classifications using radio frequency sensing modality, which is effective, safe, non-intrusive, operates in all lighting conditions, and most importantly, preserves privacy. Successful recognitions of activities of daily living (ADL) such as walking, kneeling, sitting, standing, bending, and falling have been reported in the literature \cite{van2008feature,kim2015human,mobasseri2009time,gurbuz2016micro,seifertsubspace,erol2018automatic,kim2016human,seyfiouglu2018deep,jokanovic2016radar,le2018human,chen2018personnel}.
 Human motion classification has been examined based on handcrafted features that relate to human motion kinematics \cite{van2008feature,kim2015human,mobasseri2009time,gurbuz2016micro}, and other approaches that are data driven and include low-dimension representations \cite{
 seifertsubspace}, frequency-warped cepstral analysis \cite{erol2018automatic}, and neural networks \cite{le2018human,kim2016human,seyfiouglu2018deep,jokanovic2016radar,chen2018personnel}.\par
Propelled by successes in discriminating between different human activities, radar has been recently employed for automatic hand gesture recognition for interactive intelligent devices \cite{li2018sparsity,kim2016hand,wang2016interacting,skaria2019hand,zhang2016dynamic,maminzz}. This recognition proves important in contactless close-range hand-held or arm-worn devices, such as cell phones and watches. The most recent project on hand gesture recognition, Soli, by Google for touchless interactions with radar embedded in a rest band is a testament of this emerging technology \cite{wang2016interacting}. In general, automatic hand or arm gesture recognition, through the use of radio frequency (RF) sensors, is important to smart environment. It is poised to make homes more user friendly and most efficient by identifying different motions for controlling instrument and household appliances. The same technology can also greatly benefit the physically challenged who might be wheelchair confined or bed-ridden patients. The goal is then to enable these individuals to be self-supported and independently functioning.\par
Arm motions assume different kinematics than those of hands especially in terms of speed and time extent. Compared to hand gesture, arm gesture recognition can be more suitable for contactless man-machine interaction with longer range separation, e.g., the case of commanding appliances, like TV, from a distant coach. The larger radar cross-section of the arms, vis-a-vis hands, permits more remote interactive positions in an indoor setting. Further, the ability of using hand gestures for device control can sometimes be limited by cognitive impairments such the Parkinson disease. In this case, arm motions can be more robust to strong hand tremor and shaking.\par
Classification approaches used for ADL are typically based on computing the spectrograms of the radar signal returns which reveal the MD of the moving targets. The same approaches can be readily applied for recognition of arm motions. However, there is an apparent difference between the MD signatures of arm motions and those associated with motion activities that involve the whole human body. Depending on the experiment setup and radar data collection specs, MD signatures of arm motions are typically simple, limited to short time durations and small frequency bandwidth, and have confined power concentrations in the time-frequency domain. Further, arm gesture MD signature is rather contiguous and does not comprise isolated energy regions in the time-frequency domain as the case with most reported hand gestures. On the other hand, the MD signatures of body motions are intricate, of multi-components, and strongly influenced by the torso. They span relatively longer time periods and assume higher Doppler frequencies.\par
In this paper, we present a simple but effective method to discriminate dynamic arm motions using a MD radar sensor. We present a classification approach that utilizes the positive and negative envelopes of the spectrogram MD signatures. It is shown that these two envelopes are shown to implicitly capture, for each motion, the intrinsic and salient features of corresponding Doppler signal behavior as well as the degree of signal power occupancy over the joint time and frequency variables. We compare the proposed approach with four different classification methods, namely, the PCA-based method \cite{seifertsubspace,Sahoo2019}, the empirical feature extraction method \cite{zhang2016dynamic}, the sparse reconstruction-based method \cite{li2018sparsity} and the CNN-based method \cite{kim2016hand,skaria2019hand}. Based on the experimental data collected and the arm motions considered, we demonstrate that the proposed approach outperforms the above methods, and achieve a classification accuracy higher than 97\%.\par
Since both arms are involved in every gesture and move together either in the same or opposite directions for all suggested motions, then unlike hand motions, there is no information gleaned from angular resolution that would help in improving classifications \cite{chen2019dynamic}. Previous techniques for RF-based arm recognition include the work by Sun \textit{et al.} \cite{sun2018gesture} who used five handcrafted MD features and $k$-NN classifier to recognize seven arm gestures measured by a frequency modulated continuous wave (FMCW) radar. In \cite{lou2018gesture}, the distance between two arm gesture signals is obtained based on the improved Dynamic Time Warping proximity matching method, which is then regarded as a distance metric in the $k$-NN classifier to distinguish five arm gestures.\par
The remainder of this paper is organized as follows. In Section II, we discuss the MD signature envelope extraction method and the proposed classification technique. Section III describes the radar data collection and pre-processing of arm motions. In Section IV, arm motion similarity based on canonical correlations is proposed to decide on proper arm motions from the classification perspective. Section V gives the experimental results using real data measurements for different classification approaches based on handcrafted features and data-driven feature extractions. Section VI is the conclusion.
\section{Methodology}
\subsection{Radar MD Signature Representation}
\subsubsection{Time-frequency Representations}
Arm motions generate non-stationary radar back-scattering signals, which are typically analyzed by time-frequency representation (TFR) methods \cite{Rao2003Time}. TFR reveals the signal local frequency behavior in the joint-variable domain referred to as the MD signature. A commonly used technique for TFRs is the spectrogram. For a discrete-time signal \(s(n)\) of length \(N\), the spectrogram can be obtained by taking the short-time Fourier transform (STFT) of the data and computing the magnitude square \cite{Allen1982Applications}, 
\begin{equation}\label{stft}
              S\left( {n,k} \right) = {\left| {\sum\limits_{m = 0}^{L - 1} {s(n + m)h(m){e^{ - j2\pi \frac{{mk}}{N}}}} } \right|^2}\
\end{equation}
where \(n=1,\cdots,N\) is the time index, $k=1,\cdots\,K$ is the discrete frequency index, and $L$ is the length of the window function $h(\cdot)$. It is noted that if the MD signal can be modeled as a sum of frequency modulated signals, then the signal parameters can be estimated using maximum likelihood techniques \cite{setlur2006analysis}. However, the MD signal of the arm motion does not conform to this model and, as such, spectrograms are used for feature extractions, and without assuming any model for feature behaviors \cite{cirillo2008parameter}.
\subsubsection{Power Burst Curve (PBC)}
The onset and offset times of each motion can be determined by monitoring the PBC \cite{erol2017range,amin2019rf}, which measures the signal energy in the spectrogram within specific frequency bands. That is,
\begin{equation}\label{pbc}
S(n) = \sum\limits_{{k_1} = {K_{N1}}}^{{K_{N2}}} {{{\left| {S(n,{k_1})} \right|}^2}}  + \sum\limits_{{k_1} = {K_{P1}}}^{{K_{P2}}} {{{\left| {S(n,{k_1})} \right|}^2}} \
\end{equation}
The negative frequency indices $K_{N1}$ and $K_{N2}$ are set to $-500 Hz$ to $-20 Hz$, whereas the indices for positive frequencies are $K_{P1}=20 Hz$ and $K_{P2}=500 Hz$. The frequency band around the zero Doppler bin between $-20 Hz$ and $20 Hz$ affects the accuracy of the result, and therefore is not considered.\par
A moving average filter is applied to smooth the original PBC curve. The filtered PBC is denoted as $S_{f}(n)$. The threshold, $T$, determines the beginning and the end of each motion and is computed by
\begin{equation}\label{thre}
T = {S_{f\min }} + \alpha  \cdot \left( {{S_{f\max }} - {S_{f\min }}} \right)\
\end{equation}
where $\alpha $ depends on the noise floor and is empirically chosen from $[0.01,0.2]$. In our work, $\alpha $ is set to 0.1, which means 10\% over the minima. The onset time of each motion is determined as the time index at which the filtered PBC exceeds the threshold, whereas the offset time corresponds to the time index at which the filtered PBC falls below the threshold.\par
\subsection{Extraction of the MD Signature Envelopes}
We select features specific to the nominal arm motion local frequency behavior and power concentrations. These features are the positive and negative frequency envelopes in the spectrograms. The envelopes represent the maximum instantaneous frequencies. They attempt to capture, among other things, the maximum positive and negative frequencies, time-duration of the arm motion event and its bandwidth, the relative portion of the motion towards and away from the radar. In this respect, the envelopes can accurately characterize different arm motions. They can be determined by an energy-based thresholding algorithm \cite{maminzz,erol2017range}. First, the effective bandwidth of each motion is computed. This defines the maximum positive and negative Doppler frequencies. Second, the spectrogram is divided into positive frequency and negative frequency parts. The corresponding energies of the two parts, denoted as ${E_U}(n)$ and ${E_L}(n)$, are computed separately as,
 \begin{equation}\label{uenergy}
              {E_U}\left( n \right) = \sum\limits_{k = 1}^{\frac{K}{2} } {S{{\left( {n,k} \right)}^2}} ,  
               {E_L}\left( n \right) = \sum\limits_{k = \frac{K}{2}+1}^{K} {S{{\left( {n,k} \right)}^2}} 
\end{equation}
These energies are then scaled to define the respective thresholds, ${T_U}$ and ${T_L}$,
\begin{equation}\label{uthreshold}
              {T_U}(n) = {E_U}(n)\cdot {\sigma_U}, {T_L}(n) = {E_L}(n)\cdot {\sigma_L}
\end{equation}
where ${\sigma_U}$ and ${\sigma_L}$ represent the scale factors, both are less than 1. These scalars can be chosen empirically, but an effective way for their selections is to maintain the ratio of the energy to the threshold values constant over all time samples. This constant ratio can be found by time locating the maximum positive Doppler frequency and computing the corresponding energy at this location. Once the threshold is computed, the positive frequency envelope is then provided by locating the Doppler frequency at each time instant for which the spectrogram assumes the first higher or equal value to the threshold. This frequency, in essence, represents the effective maximum instantaneous Doppler frequency. Similar procedure can be followed for the negative frequency envelope.  The positive frequency envelope, ${e_U}(n)$, and negative frequency envelope, $e_L(n)$, are concatenated to form the feature vector $e=[{e_U},{e_L}]$.
\subsection{Proposed classification method}
The method considered is motivated by the contiguity of the arm motion MD signatures and by our investigations to data-driven feature extractions using PCA applied to the spectrograms. It is found that the PCA-based classification results, irrespective of the employed classifier, insignificantly change when considering only the maximum instantaneous Doppler frequencies rather than using the entire spectrogram. Only focusing on the maximum instantaneous frequencies is accomplished by starting with the spectrogram and then assigning unit values to the MD signature envelope, with the rest of the spectrogram values set to zero. The performance similarity between using the spectrogram vs. the envelopes is also exhibited when employing CNN. Motivated by such findings, we consider the MD signature envelopes as the sole features and proceed to classify arm motions using the NN classifier, without performing PCA. In so doing, we considerably reduce memory and computations, with performance comparable to that of PCA and CNN classification. Just as an example, a PCA acting on the entire spectrogram of dimension $N$-by-$N $would process vectors of dimensions $N^2$, whereas the feature envelope vector is only of dimension $2N$, if both the positive and negative behaviors are considered. The fact that the NN classifier performs well when operating on the MD envelopes suggests very important and interesting property of the arm gesture motions. Basically, it is the envelope values rather than the envelope particular shapes that guide classification performance. In essence, irrespective of the closeness or the distance measure employed, shuffling the envelope of one motion, i.e., randomly rearranging the envelope values over the time axis, does not alter the results.
\section{Arm Motion Experiments}
\begin{figure}[htbp]
\centering
\includegraphics[width=0.45\textwidth]{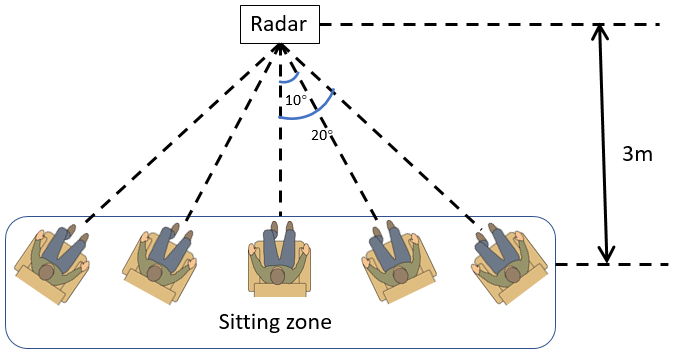}
\caption{Illustration of experiment setup}
\label{Sit}
\end{figure}

The data analyzed in this paper was collected in the Radar Imaging Lab at the Center for Advanced Communications, Villanova University. The system in the experiments utilizes one K-band portable radar sensor from the Ancortek company with one transmitter and one receiver. It generates a continuous wave (CW) with the carrier frequency 25 GHz and the sampling rate is 12.8 kHz. The radar was placed at the edge of a table. The arm motions were performed at approximately three meters away from radar in a sitting position of the participants. The body remained fixed as much as possible during the experiments. In order to mimic typical people behavior, the arms are always resting down at a table or arm chair level at the initiation and conclusion of each arm motion. In the experiments, we choose five different orientation angles, $0,\pm 10^\circ,\pm 20^\circ$, as shown in Fig. \ref{Sit}, with the person always facing the radar. Different speeds of the arm motion are also considered. It is noted that the elderly are likely to perform arm motions with a slower pace than the young, so the experiments contain both normal and slow arm motions. The latter one is about 30\% slower than the former.
\par

\begin{figure}[htbp]
\begin{minipage}[b]{0.1\textwidth}
\includegraphics[width=1\textwidth]{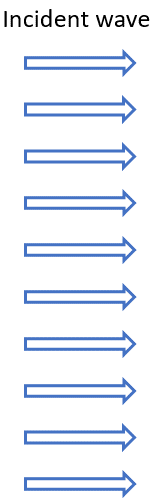} 
\vspace{0.1cm}
\end{minipage}
\begin{minipage}[b]{0.38\textwidth}
\includegraphics[width=1\textwidth]{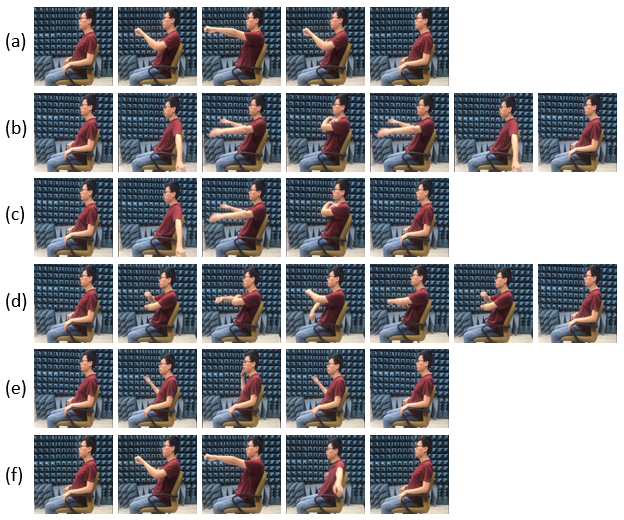}\\
\end{minipage}
  \caption{ Illustrations of 6 different arm motions. (a) Pushing arms and pulling back, (b) Crossing arms and opening, (c) Crossing arms, (d) Rolling arms, (e) Stop sign, (f)Pushing arms and opening.} 
\label{photo}
\end{figure}

As depicted in Fig.\ref{photo}, the following six arm motions were conducted: (a) Pushing arms and pulling back, (b) Crossing arms and opening, (c) Crossing arms, (d) Rolling arms, (e) Stop sign, and (f) Pushing arms and opening. In ``pushing," the arms moved towards the radar, whereas in ``pulling," they moved away from the radar. Both motions are relatively quick, with ``pulling" immediately following ``pushing." The motion of ``crossing arms" describes crossing the arms from a wide stretch. Six people participated in the experiment. Each arm motion was recorded over 40 seconds to generate one data segment. The recording was repeated 4 times, containing slow and normal motions at each angle. Each data segment contained 12 or 13 individual arm motions, and a 5 second time window is applied to capture the individual motions according to the onset and offset time determined by the PBC. As such, repetitive motions and the associated duty cycles were not considered as features and were not part of the classifications. In total, 1913 segments of data for six arm motions were generated. The most discriminative arm motion can be used as an ``attention'' motion for signaling the radar to begin, as well as to end, paying attention to the follow on arm motions. That is, without the ``attention'' motion, the radar remains passive with no interactions with human. Among all arm motions, ``Pushing arms and pulling back'' and ``Pushing and open arms'' assume the highest accuracy. However, the former motion can be confused with common arm motions such as reaching for a cup or glasses on table. Thus, ``Pushing and open arms'' is chosen as the ``attention'' motion.\par

\begin{figure}[htbp]
\begin{subfigure}[b]{0.5\linewidth} 
\centering\includegraphics[width=1\linewidth]{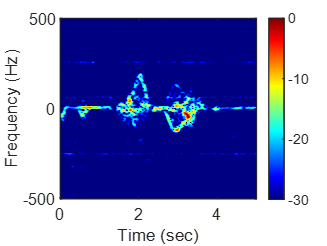} 
\captionsetup{justification=centering}
\caption{} 
\end{subfigure}\hfill
\begin{subfigure}[b]{0.5\linewidth} 
\centering\includegraphics[width=1\linewidth]{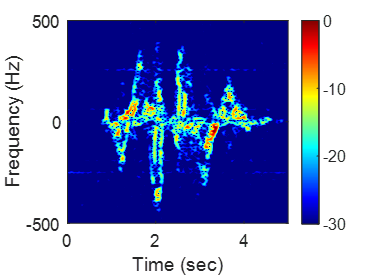} 
\captionsetup{justification=centering}
\caption{} 
\end{subfigure}\vspace{10pt}
\begin{subfigure}[b]{0.5\linewidth} 
\centering\includegraphics[width=1\linewidth]{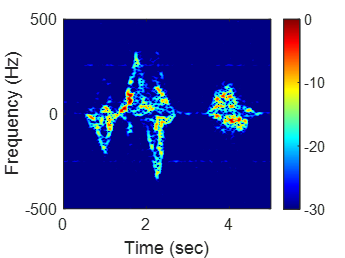} 
\captionsetup{justification=centering}
\caption{} 
\end{subfigure}\hfill
\begin{subfigure}[b]{0.5\linewidth} 
\centering\includegraphics[width=1\linewidth]{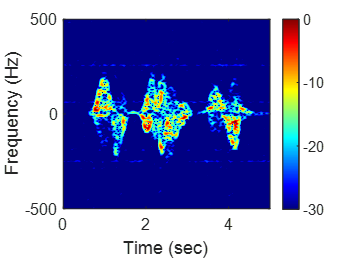} 
\captionsetup{justification=centering}
\caption{} 
\end{subfigure}\vspace{10pt}
\begin{subfigure}[b]{0.5\linewidth} 
\centering\includegraphics[width=1\linewidth]{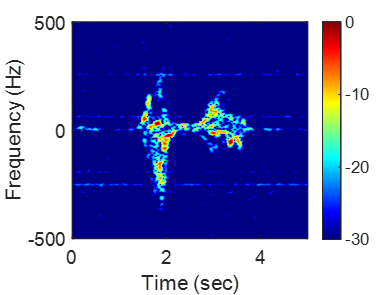} 
\captionsetup{justification=centering}
\caption{} 
\end{subfigure}\hfill
\begin{subfigure}[b]{0.5\linewidth} 
\centering\includegraphics[width=1\linewidth]{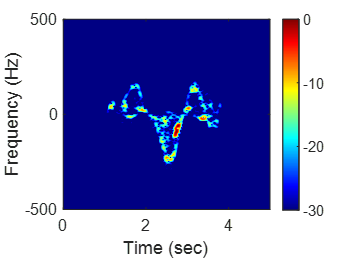} 
\captionsetup{justification=centering}
\caption{} 
\end{subfigure}
  \caption{ Spectrograms of 6 different arm motions. (a) Pushing arms and pulling back, (b) Crossing arms and opening, (c) Crossing arms, (d) Rolling arms, (e) Stop sign, (f)Pushing arms and opening.} 
\label{spectrograms}
\end{figure}

\begin{figure}[t]
\begin{subfigure}[b]{0.5\linewidth} 
\centering\includegraphics[width=1\linewidth]{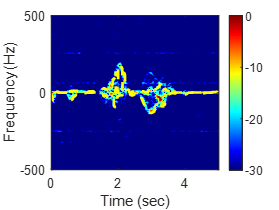} 
\captionsetup{justification=centering}
\caption{} 
\end{subfigure}\hfill
\begin{subfigure}[b]{0.5\linewidth} 
\centering\includegraphics[width=1\linewidth]{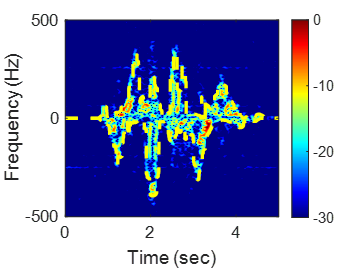} 
\captionsetup{justification=centering}
\caption{} 
\end{subfigure}\vspace{10pt}
\begin{subfigure}[b]{0.5\linewidth} 
\centering\includegraphics[width=1\linewidth]{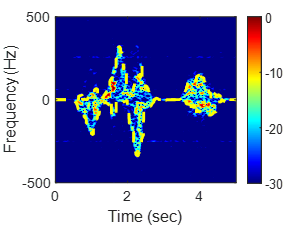} 
\captionsetup{justification=centering}
\caption{} 
\end{subfigure}\hfill
\begin{subfigure}[b]{0.5\linewidth} 
\centering\includegraphics[width=1\linewidth]{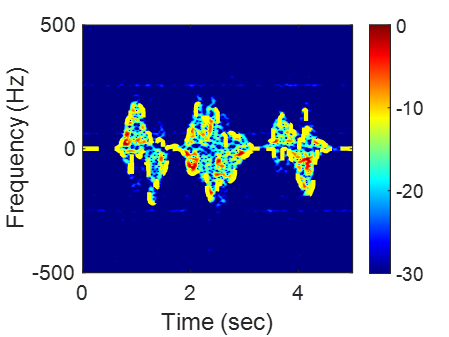} 
\captionsetup{justification=centering}
\caption{} 
\end{subfigure}\vspace{10pt}
\begin{subfigure}[b]{0.5\linewidth} 
\centering\includegraphics[width=1\linewidth]{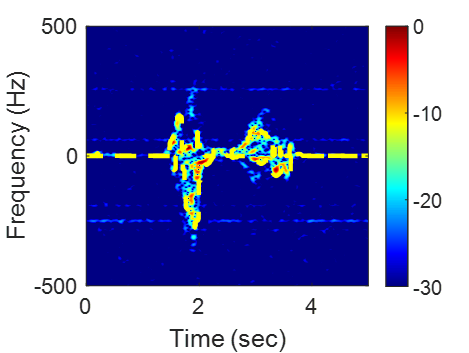} 
\captionsetup{justification=centering}
\caption{} 
\end{subfigure}\hfill
\begin{subfigure}[b]{0.5\linewidth} 
\centering\includegraphics[width=1\linewidth]{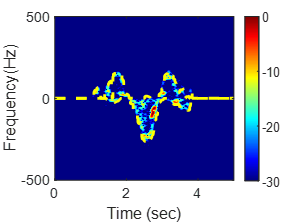} 
\captionsetup{justification=centering}
\caption{} 
\end{subfigure}
  \caption{ Spectrograms and corresponding envelopes of 6 different arm motions. (a) Pushing arms and pulling back, (b) Crossing arms and opening, (c) Crossing arms, (d) Rolling arms, (e) Stop sign, (f)Pushing arms and opening.} 
\label{envelopes}
\end{figure}

Fig. \ref{spectrograms} shows examples of spectrograms for different arm motions with normal speed at zero angle. The employed sliding window $h(\cdot)$ is rectangular with length $L=$2048 (0.16 $s$), and $K$ is set to 4096. The envelopes are extracted and plotted in Fig.\ref{envelopes}. It is clear that the envelopes can well enclose the local power distributions. It is also evident that the MD characteristics of the spectrograms are in agreement and consistent with each arm motion kinematics. For example, in ``Pushing arms and pulling back,'' the arms push forward directly which generates positive frequencies, whereas the ``pulling'' phase has negative frequencies. The arm motion, ``Crossing arms and opening,'' can be decomposed into two phases. In the ``crossing'' phase, the arms move closer to the radar at the beginning which causes the positive frequencies, then move away from the radar which induces the negative frequencies. The ``open'' phase is the opposite motion of ``crossing'' phase, which also produces the positive frequencies first and then negative frequencies. The motion ``Crossing arms'' only contains the first phase of the motion ``Crossing arms and opening,'' and has the same respective MD signature. The two arms of ``Rolling arms'' perform exactly the opposite movements, as one arm moves forward along a circular trajectory, the other moves backwards. So, the MD has simultaneously positive and negative frequencies. In one motion cycle, the right arm experience three phases, moving forward, moving backward and moving forward again. The left arm always performs the opposite motion to the right arm. For the motion, ``Stop sign,'' the arm moves backwards which only causes negative frequencies. The last arm motion, ``Pushing arms and opening '' includes the pushing, which has positive frequencies, and the opening, which has negative frequencies.\par

\begin{figure}[htp]
\centering\begin{subfigure}[b]{0.5\linewidth} 
\centering\includegraphics[width=1\linewidth]{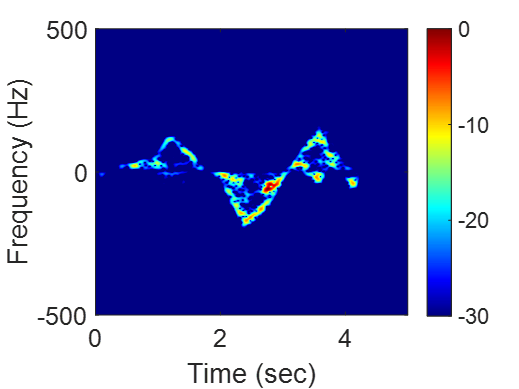} 
\captionsetup{justification=centering}
\caption{} 
\end{subfigure}\hfill
\begin{subfigure}[b]{0.5\linewidth} 
\centering\includegraphics[width=1\linewidth]{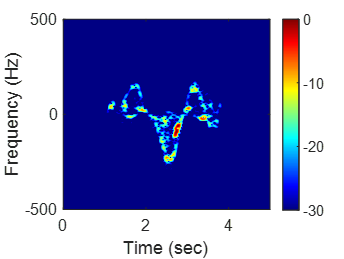} 
\captionsetup{justification=centering}
\caption{} 
\end{subfigure}
  \caption{ The ``attention'' motion with different velocities at $0^\circ$. (a) Slow motion, (b) normal motion.} 
\label{velocity}
\end{figure}

\begin{figure}[htp]
\centering\begin{subfigure}[b]{0.5\linewidth} 
\centering\includegraphics[width=1\linewidth]{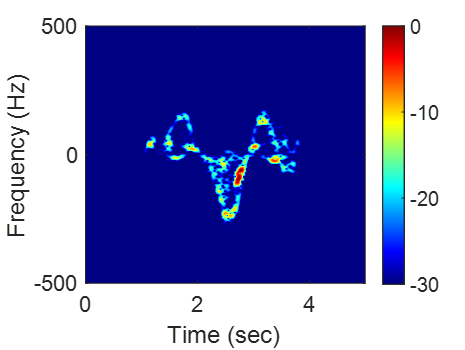} 
\captionsetup{justification=centering}
\caption{} 
\end{subfigure}\hfill
\begin{subfigure}[b]{0.5\linewidth} 
\centering\includegraphics[width=1\linewidth]{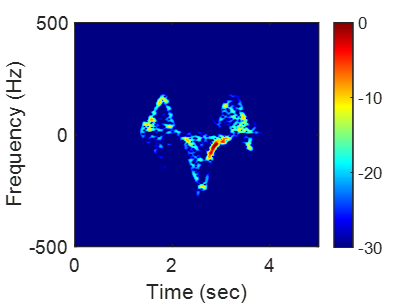} 
\captionsetup{justification=centering}
\caption{} 
\end{subfigure}\vspace{10pt}

\begin{subfigure}[b]{\linewidth} 
\centering\includegraphics[width=.5\linewidth]{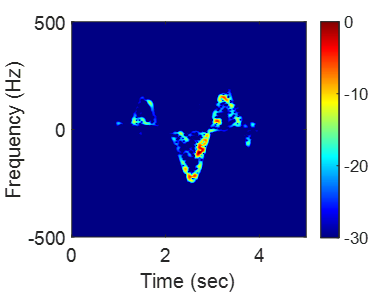} 
\captionsetup{justification=centering}
\caption{} 
\end{subfigure} 
  \caption{ The ``attention'' motion with normal speed at different orientation angles. (a) The ``attention'' motion at $0^\circ$, (b) The ``attention'' motion at $10^\circ$, (c) The ``attention'' motion at $20^\circ$.} 
\label{angle}
\end{figure}

Fig. \ref{velocity} is an example of the ``attention'' motion with different velocities at $0^\circ$. The time period of the normal motion is shorter than that of the slow motion, and the speed is faster which causes higher Doppler frequencies. The main characteristics and behaviors, however, remain unchanged. Fig. \ref{angle} shows the ``attention'' motion with the normal speed at different orientation angles. As the angle increases, the energy becomes lower owing to the $dB$ drop in the antenna beam.

\par
\section{Arm Motion Similarity Measures and Canonical Correlations}
Observing the spectrograms in Fig. \ref{spectrograms}, it is noticeable that the signatures of different motions are distinguishable. To mathematically describe these dissimilarities, we consider the Canonical correlation measure \cite{jokanovic2017suitability}. In this case, the spectrograms are converted to gray-scale images with the size $100\times 100$, and then vectorized with the size $1\times 10000$. Define matrix $X$ that contains $M$ vectorized images ${x_i}, i=1,\cdots,M$ of a specific arm motion,
\begin{equation}\label{classmatrix}
            X = [{x_1}|{x_2}| \cdots |{x_M}]
\end{equation}
The $d$-dimensional subspace of a specific arm motion can be obtained by performing the singular value decomposition (SVD) on $X$ \cite{jolliffe2011principal}. Suppose $\Phi_1$ and $\Phi_2$ are two $d$-dimensional linear subspaces, the canonical correlations of the two subspaces are the cosines of principal angles, and are defined as \cite{kim2007discriminative},
\begin{equation}\label{angle1}
            \cos {\theta _i} = \mathop {\max }\limits_{{u_i} \in {\Phi _1}} \mathop {\max }\limits_{{v_i} \in {\Phi _2}} u_i^T{v_i}
\end{equation}
subject to $||u||=||v||=1, {u_i}^Tu_j={v_i}^Tv_j=0, i\ne j$. Let $U$ and $V$ denote the orthogonal bases for the two subspaces, $\Phi_1$ and $\Phi_2$. The SVD of $U^TV$ is,
\begin{equation}\label{svd}
            U^TV=P\Lambda Q
\end{equation}
The canonical correlations are the singular values $\Lambda$, i.e., $\cos (\theta_i)=\lambda_i, i=1,\cdots,d$. The minimum angle is used to measure the closeness of two subspaces. Table \ref{canonical} shows the canonical correlations coefficients of the motion considered, from which we can clearly deduce the dissimilarities between the different arm motions. All the coefficients assume small values less than 0.65 which indicates low resemblances. Hence, the six arm motions are suitable candidates for classification. It is important to note that other similarity measures \cite{mitchell2010image} can be applied, in lieu of the canonical correlation. However, we found the canonical correlation most consistent with the visual similarities. 

\begin{table}[htbp]
\centering
\caption{ \sc Canonical Correlations Coefficients}
\begin{tabular}{cccccc}
\hline\hline
 
     & b       & c       & d       & e       & f  \\ \hline
a &0.41 &0.39& 0.49  & 0.48 & 0.56   \\            
b & 0  & 0.41 &0.29 & 0.28 &0.27   \\   
c & 0& 0& 0.61 &0.51  & 0.32   \\ 
d & 0  & 0 &0  &0.43 &0.44  \\
e &0  & 0 &0  &0  & 0.44 \\   \hline\hline
\end{tabular}
\label{canonical}
\end{table}

\section{Experimental Results}
In this section, all 1913 data segments are used to validate the proposed method where 70\% of the segments are used for training and 30\% for testing. The classification results are obtained by 100 Monte Carlo trials. Four different automatic arm motion approaches are compared with the proposed method. These are: 1) the PCA-based method \cite{seifertsubspace,Sahoo2019}; 2) the empirical feature extraction method \cite{zhang2016dynamic}; 3) the sparse reconstruction-based method \cite{li2018sparsity}; 4) the CNN-based method \cite{kim2016hand,skaria2019hand}.
\subsection{PCA-based methods}
Given $M$ spectrograms as training samples $X_i, i=1,\dots,M$, each spectrogram is first resized to an $N$-by-$N$ image and then vectorized as ${x_i} = vec\{ {X_i^T}\}$ whose length is $Q\times 1$, with $Q=N^2$ representing the total number of the pixels in the image.\par
\textit{ PCA of Spectrograms:} For the PCA applied to the spectrograms, each sample represents a spectrogram image of $100\times 100$ pixels. The number of principal components is determined by the dominant eigenvalues, and found to be equal to 30 for best results. Table \ref{pcacon} is the corresponding confusion matrix showing that this PCA method can achieve an overall accuracy of 95.91\%.  \par

\begin{figure}[htbp]
\centering
\includegraphics[width=0.4\textwidth]{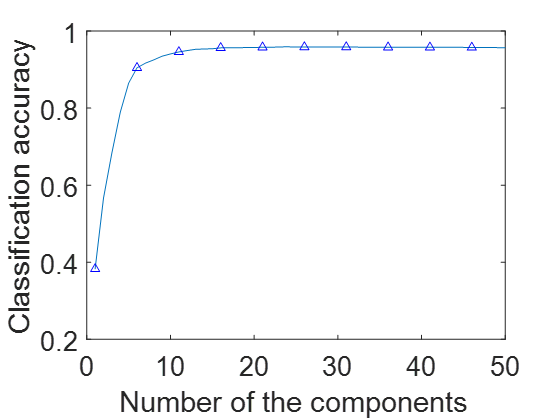}
\caption{Performance of PCA with different number of principal components.}
\label{PCA}
\end{figure}
\begin{table}[htbp]
\centering
\caption{  \sc Confusion Matrix Yielded by PCA-based Method with $d=30$ for the Entire Spectrogram}
\begin{tabular}{ccccccc}
\hline\hline
  & a       & b       & c       & d       & e       & f\\ \hline
a & 97.15\% & 0.02\%  & 0.40\%  & 0.24\%  & 0.81\%  & 1.38\%\\ 
b & 0.65\%  & 92.16\% & 3.19\% & 0.34\%  & 2.08\%  & 1.58\%\\  
c & 1.31\% & 1.77\% & 92.23\% & 0.52\%  & 2.30\%  & 1.87\%\\   
d & 2.68\%  & 0.48\% & 0.76\%  & 95.05\% & 0.65\% & 0.38\% \\     
e & 1.21\%  & 0.01\%  & 0.27\%  & 0.17\%  & 97.64\% & 0.70\% \\ 
f& 1.65\% & 0.04\%& 0.11\% & 0.05\%  & 0.26\% & 97.89\% \\ \hline  \hline
\end{tabular}
\label{pcacon}
\end{table}

\begin{figure}[htbp]
\begin{subfigure}[b]{0.33\linewidth} 
\centering\includegraphics[width=1\linewidth]{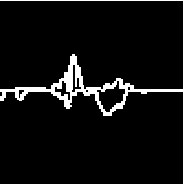} 
\captionsetup{justification=centering}
\caption{} 
\end{subfigure}\hfill
\begin{subfigure}[b]{0.33\linewidth} 
\centering\includegraphics[width=1\linewidth]{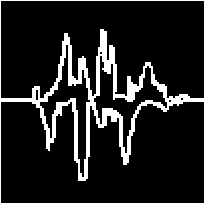} 
\captionsetup{justification=centering}
\caption{} 
\end{subfigure}\hfill
\begin{subfigure}[b]{0.33\linewidth} 
\centering\includegraphics[width=1\linewidth]{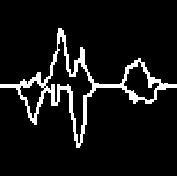} 
\captionsetup{justification=centering}
\caption{} 
\end{subfigure}\vspace{10pt}
\begin{subfigure}[b]{0.33\linewidth} 
\centering\includegraphics[width=1\linewidth]{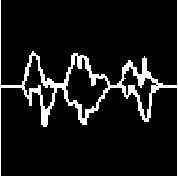} 
\captionsetup{justification=centering}
\caption{} 
\end{subfigure}\hfill
\begin{subfigure}[b]{0.33\linewidth} 
\centering\includegraphics[width=1\linewidth]{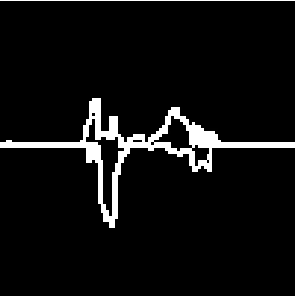} 
\captionsetup{justification=centering}
\caption{} 
\end{subfigure}\hfill
\begin{subfigure}[b]{0.33\linewidth} 
\centering\includegraphics[width=1\linewidth]{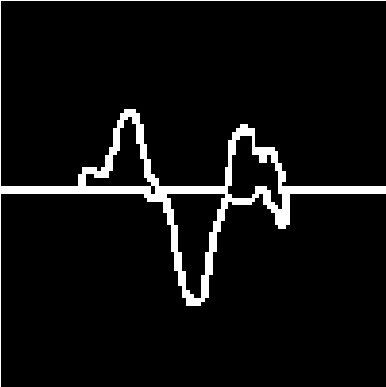} 
\captionsetup{justification=centering}
\caption{} 
\end{subfigure}
  \caption{ Envelope images of 6 different arm motions. (a) Pushing arms and pulling back, (b) Crossing arms and opening, (c) Crossing arms, (d) Rolling arms, (e) Stop sign, (f)Pushing arms and opening.}
\label{envimage}
\end{figure}

\textit{ PCA of Spectrograms with Only MD Signature Envelopes:}  In order to understand the role of the envelopes in the spectrogram image, we set the image values to one at the locations of the maximum instantaneous Doppler frequencies, while the rest of the image is set to zero. Different from the original images, the new spectrogram image only informs us with the time locations of the highest frequencies. The spectrograms with only envelope locations are shown in Fig.\ref{envimage}. Following the same PCA steps, but applied to the modified spectrogram image, surprisingly, the classification achieves 95.89\% accuracy. This is almost the same performance as the PCA operating on the original spectrogram images. The corresponding confusion matrix is shown in Table \ref{envimgpcacon}. This performance similarity not only underscores the importance of preserving the maximum instantaneous frequencies but also highlights the relatively weak role of other MD frequencies in motion discrimination.\par

\begin{table}[htbp]
\centering
\caption{ \sc Confusion Matrix Yielded by PCA-based Method with $d=30$ for the Spectrograms with Only Envelopes}
\begin{tabular}{ccccccc}
\hline\hline
  & a       & b       & c       & d       & e       & f\\ \hline
a & 97.33\% & 0  & 0.21\%  & 0.37\%  & 1.54\% & 0.55\%\\ 
b & 0.04\%  & 93.87\% & 3.57\% & 0.72\%  & 0.38\%  & 1.42\%\\  
c & 0.81\% & 0.38\% & 92.70\% & 1.06\%  & 3.86\%  & 1.19\%\\   
d & 1.12\%  & 0.03\% & 0.49\%  & 97.43\% & 0.58\% & 0.35\% \\     
e & 1.05\%  & 0  & 4.02\%  & 0.50\%  & 93.63\% & 0.80\% \\ 
f & 0.79\% & 0 & 0.07\% & 0.06\%  & 1.06\% & 98.02\% \\ \hline  \hline
\end{tabular}
\label{envimgpcacon}
\end{table}

\textit{ PCA of Only Vectorized Envelopes:} Instead of performing the PCA on the images, the vectorized envelope $e=[{e_U},{e_L}]$ itself can be regarded as a high dimensional data, so the PCA can also be directly applied to the envelope vector. The size of the envelope vectors is only $2N$. Table \ref{envpcacon} shows the confusion matrix by performing PCA directly on the vectorized envelopes with overall accuracy of 96.14 \%.\par
\begin{table}[htbp]
\centering
\caption{ \sc Confusion Matrix Yielded by PCA-based Method with $d=30$ for the Vectorized Envelopes}
\begin{tabular}{ccccccc}
\hline\hline
  & a       & b       & c       & d       & e       & f\\ \hline
a & 96.71\% & 1.52\%  & 0.13\%  & 0.23\%  & 1.59\% & 0.02\%\\ 
b & 2.12\%  & 93.80\% & 3.26\% & 0.35\%  & 0  & 0.47\%\\  
c & 0.21\% & 2.46\% & 94.31\% & 1.76\%  & 0.83\%  & 0.43\%\\   
d & 0.86\%  & 0 & 2.18\%  & 95.44\% & 1.52\% & 0 \\     
e & 1.96\%  & 0  & 1.27\%  &2.41\%  & 94.21\% & 0.15\% \\ 
f & 0.14\% & 0.07\% & 0.11\% & 0.10\%  & 0.21\% & 99.37\% \\ \hline  \hline
\end{tabular}
\label{envpcacon}
\end{table}

It is worth noting that the PCA applied to the spectrograms deals with the images of the size $N\times N$, whereas the dimension of the envelopes is only $2N$. From the results of the above three PCA-based methods, a strong argument can be made that the MD envelopes uniquely characterize the corresponding motions, which suggests that they can be used as features without the need for PCA.

\subsection{Proposed Envelope-based Method}
We apply the NN classifier acting on the MD envelopes. Four different distance metrics are considered, namely, the Euclidean distance, the Manhattan distance \cite{wang2007improving}, the Earth Mover's distance (EMD) \cite{rubner2001earth} and the modified Hausdorff distance (MHD) \cite{dubuisson1994modified}. We also use the support vector machine (SVM) classifier for comparison. The recognition accuracy is presented in Table \ref{envacc}. It is clear that the NN classifier based on L1 distance achieves the highest accuracy. Different from other distances, the L1 distance attempts to properly account for small envelope values. The confusion matrix of the NN classifier based on the L1 distance is shown in Table \ref{envcon}, from which we can observe that motion (a) and motion (f) are most distinguishable, with an accuracy over 98\%. Since the distance measure is the sum of the differences, whether in absolute or squared values, of the corresponding elements in the test and training envelope vectors, then the values assumed by each envelope, rather than the envelope evolutionary shape, are fundamental to the classification performance. 

\begin{table}[htbp]
\centering
\caption{\sc Recognition Accuracy with Different Types of Classifier}
\begin{tabular}{c c}
\hline\hline
 & \textbf{Accuracy }\\ \hline
SVM & 83.46\% \\ 
NN-L1 & 97.17\% \\ 
NN-L2 & 96.72\% \\ 
NN-EMD & 96.78\% \\ 
NN-MHD & 96.86\% \\ \hline\hline
\end{tabular}
\label{envacc}
\end{table}

\begin{table}[htbp]
\centering
\caption{ \sc Confusion Matrix Yielded by Envelope Method Based on NN-L1 Classifier}
\begin{tabular}{ccccccc}
\hline\hline
  & a       & b       & c       & d       & e       & f  \\ \hline
a & 99.17\% &0 & 0.02\%  & 0  & 0.77\%    & 0.04\% \\
b & 0.01\%  & 94.96\% & 2.68\% & 0.54\%  &0.51\%   &1.30\% \\
c & 1.04\% & 0.26\%& 95.55\% &0.12\%  & 2.44\%  & 0.59\% \\
d & 2.79\%  & 0 & 0.22\%  &96.31\% & 0.68\% & 0\\
e & 2.64\%  & 0  & 0.69\%  & 0 & 96.67\% & 0\\ 
f & 0.63\%  & 0.01\%  & 0.09\%  & 0 &0.53\% & 98.74\%\\ \hline\hline
\end{tabular}
\label{envcon}
\end{table}

\subsection{Empirical Feature Extraction Method}
\begin{figure}[htbp]
\centering
\includegraphics[width=0.35\textwidth]{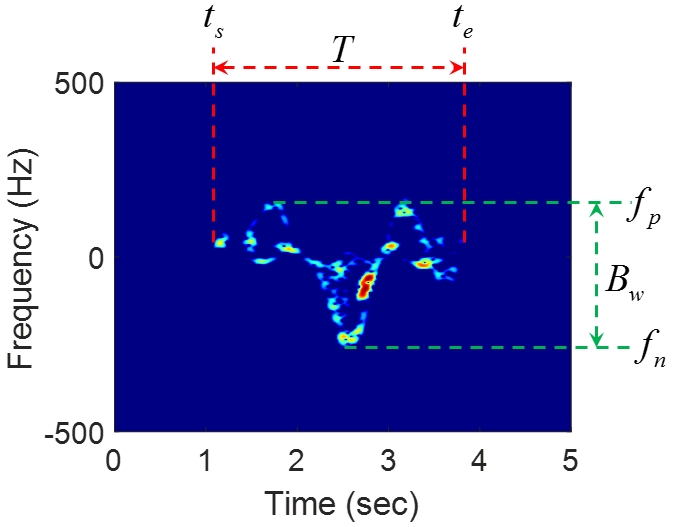}
\caption{Empirical feature extraction.}
\label{exspec}
\end{figure}

Three empirical features, as in \cite{zhang2016dynamic}, are extracted from the spectrograms to describe the arm motions, namely, the length of the event, the ratio of positive-negative frequency and the signal bandwidth. Fig. \ref{exspec} is an example showing these handcrafted features.\par

\begin{figure}[htbp]
\centering
\includegraphics[width=0.35\textwidth]{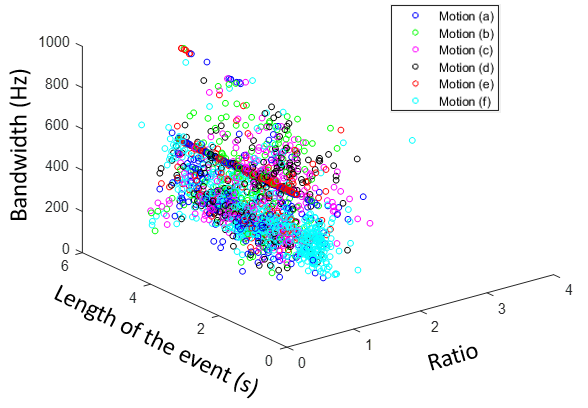}
\caption{Scatter plot of three extracted empirical features.}
\label{scatter}
\end{figure}
\textit{1) Length of the event $T$:} This describes the effective time duration to perform each arm motion,
\begin{equation}\label{timelength}
            T=t_e-t_s
\end{equation}
where $t_s$ and $t_e$ represent the onset time and the offset time of a single arm motion, respectively.
\par\textit{2) Ratio of positive-to-negative peak frequencies $R$:} This feature is obtained by finding ratio of the maximum positive frequency Doppler value, $f_p$, and maximum negative Doppler frequency value, $f_n$,
\begin{equation}\label{ratio}
            R=\left| {\frac{{{f_p}}}{{{f_n}}}} \right|
\end{equation}
where $|\cdot|$ is the absolute function.
\par\textit{3) Bandwidth $B_w$:} This is a measure of the the signal effective bandwidth,
\begin{equation}\label{bandwidth}
            B_w=|f_p|+|f_n|
\end{equation}

The scatter plot of the above features is shown in Fig. \ref{scatter}. It can be seen that the ratio of of all the motions is roughly equal to 1, the bandwidth and length of the event are also concentrated in a certain range. These motion can not be well classified by these common handcrafted features. When using NN-L1 as the classifier, the recognition accuracy based on these features is only 37.13\% with the confusion matrix shown in Table \ref{efcon}.

\begin{table}[htbp]
\centering
\caption{ \sc Confusion Matrix Yielded by Empirical Feature Extraction Method}
\begin{tabular}{ccccccc}
\hline\hline
  & a       & b       & c       & d       & e       & f  \\ \hline
a &28.40\% &15.61\%& 11.44\%  & 11.58\%  & 16.01\%   & 16.96\% \\            
b & 16.21\%  & 24.46\% &14.86\% & 14.60\%  &15.64\%   &14.23\% \\   
c & 11.88\%& 14.09\%& 29.89\% &18.15\%  & 11.21\%  & 14.78\% \\ 
d & 12.47\%  & 12.13\% &18.12\%  &25.01\% &15.90\% & 16.37\%\\
e &15.55\%  & 13.59\% &10.55\%  &14.91\%  & 33.42\% & 11.98\%\\   
f &8.95\%  &6.06\% &9.34\%  & 9.37\% &6.14\% & 60.14\%\\ \hline \hline
\end{tabular}
\label{efcon}
\end{table}


\subsection{Sparsity-based Method}
\begin{figure}[htbp]
\begin{subfigure}[b]{0.5\linewidth} 
\centering\includegraphics[width=1\linewidth]{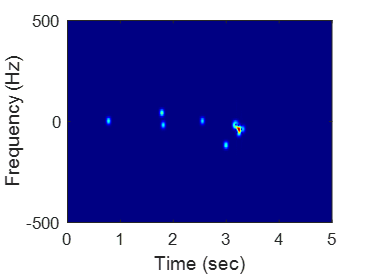} 
\captionsetup{justification=centering}
\caption{} 
\end{subfigure}\hfill
\begin{subfigure}[b]{0.5\linewidth} 
\centering\includegraphics[width=1\linewidth]{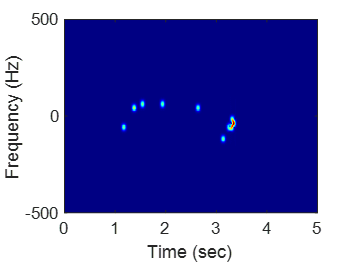} 
\captionsetup{justification=centering}
\caption{} 
\end{subfigure}\vspace{10pt}
\begin{subfigure}[b]{0.5\linewidth} 
\centering\includegraphics[width=1\linewidth]{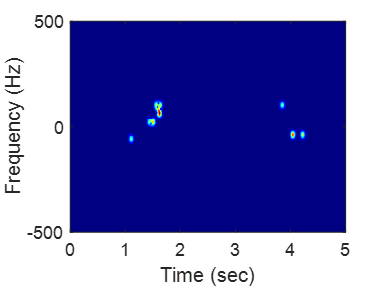} 
\captionsetup{justification=centering}
\caption{} 
\end{subfigure}\hfill
\begin{subfigure}[b]{0.5\linewidth} 
\centering\includegraphics[width=1\linewidth]{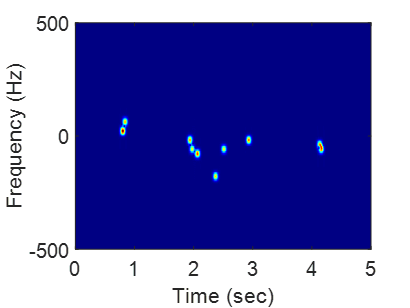} 
\captionsetup{justification=centering}
\caption{} 
\end{subfigure}\vspace{10pt}
\begin{subfigure}[b]{0.5\linewidth} 
\centering\includegraphics[width=1\linewidth]{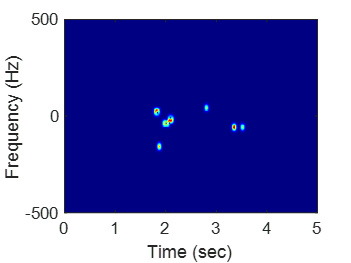} 
\captionsetup{justification=centering}
\caption{} 
\end{subfigure}\hfill
\begin{subfigure}[b]{0.5\linewidth} 
\centering\includegraphics[width=1\linewidth]{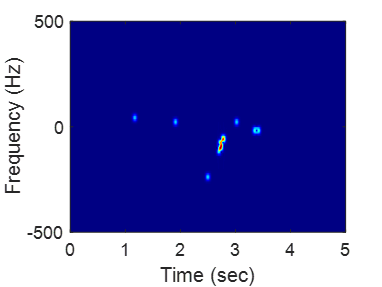} 
\captionsetup{justification=centering}
\caption{} 
\end{subfigure}
  \caption{ Spectrograms of reconstructed signals with $P = 10$.  (a) Pushing arms and pulling back, (b) Crossing arms and opening, (c) Crossing arms, (d) Rolling arms, (e) Stop sign, (f)Pushing arms and opening.} 
\label{sspectrograms}
\end{figure}

The features used for this method are the time-frequency trajectories. Details of the sparsity-based method can be found in \cite{li2018sparsity}. The trajectory consists of three parameters, namely the time-frequency position $(t_i,f_i), i=1,\cdots,P$ and the intensity $A_i$, where $P$ is the sparsity level, which is set to 10 in this paper. Hence, each sample contains 30 features. The spectrograms of reconstructed signals are plotted in Fig. \ref{sspectrograms}. In the training process, the $K$-means algorithm is used to cluster a central time-frequency trajectory. In the testing process, the NN classifier based on the modified Hausdorff distance is applied to measure the distance between the testing samples and central time-frequency trajectories. The corresponding confusion matrix is given in Table \ref{sparsecon}. The overall recognition accuracy was found to be only about 37.86\% when applied to our data. In this case, 10 sparse time-frequency positions do not properly describe the motions, especially for motion (b), motion (c) and motion (d). The reconstruction attempts to capture the strong parts of the spectrogram while ignoring the weak parts, which is evident in the arm motion reconstruction (a) and motion (f).


\begin{table}[htbp]
\centering
\caption{  \sc Confusion Matrix Yielded by Sparsity-based Method}
\begin{tabular}{ccccccc}
\hline\hline
  & a       & b       & c       & d       & e       & f  \\ \hline
a &33.45\% &13.80\%& 10.12\%  & 13.86\%  & 9.84\%   & 18.93\% \\            
b & 17.07\%  & 26.93\% &18.74\% & 14.67\%  & 8.55\%   &14.04\% \\   
c &11.52\%& 20.17\%& 26.41\% & 18.31\%  & 10.08\%  & 13.51\% \\ 
d & 14.81\%  & 12.95\% &18.98\%  &29.43\% &11.37\% &12.46\%\\
e &12.42\%  &11.14\% &12.71\%  & 14.04\%  & 32.50\% & 17.19\%\\   
f &10.69\%  &5.98\% & 6.53\%  & 7.99\% &10.03\% & 58.78\%\\ \hline \hline
\end{tabular}
\label{sparsecon}
\end{table}

\subsection{CNN-based Method}
The CNN is a widely used as an effective method in image classification. While there are many possible different CNN structures and variants one can choose from, we opt to follow a similar architecture to the one used in \cite{kim2016hand,skaria2019hand}. The input data are the spectrogram images which are the same as in the PCA method in Section V-A. We tried one, two and three CNN layers. The number of the filters for each layer is chosen empirically, and the CNN has 8 filters in the first layer, 16 filters in the second layer and 32 filters in the third layer. The filter size is chosen as 3$\times$3, 5$\times$5, or 7$\times$7, and it is the same in each layer within the CNN structure. The arm recognition results with different filter size and number of layers are shown in Table. \ref{racnn}. The highest accuracy of 96.63\%  is achieved with 3 layers, and filter size $5\times5$. The confusion matrix is given in Table \ref{dcnncon}.\par
Just as in the case of PCA-based classification of Section V-A, we used the envelope images as input to CNN. The same CNN structure of 3 layers and filter size $5\times5$ is used. In this case, arm motions can be classified with 95.16\% accuracy. This represents a small drop from the case of using the original spectrograms. It again indicates that the envelopes are important features and have large contributions to arm motion recognition.

\begin{table}[htbp]
\centering
\caption{\sc Recognition Accuracy with Different Layers and Filter Size}
\begin{tabular}{cccc}
\hline\hline
  & $3\times3$       & $5\times5$       & $7\times7$     \\ \hline
One layer &94.72\% &94.67\%&94.98\% \\            
Two layers &95\% & 95.30\% &95.70\%  \\   
Three layers &95.70\%&96.63\%& 96.60\%  \\ \hline \hline
\end{tabular}
\label{racnn}
\end{table}

\begin{table}[htbp]
\centering
\caption{ \sc Confusion Matrix Yielded by CNN-based Method}
\begin{tabular}{ccccccc}
\hline\hline
  & a       & b       & c       & d       & e       & f  \\ \hline
a &97.13\% &0.42\%&0.16\%  & 0.67\% & 1.07\%  & 0.55\%\\            
b &0 & 96.45\% &2.26\% & 0.40\% & 0.09\%  &0.80\% \\   
c &0.60\%& 1.11\%& 93.74\% & 2.58\%  &1.19\%  &0.78\% \\ 
d &0.70\%  & 0.40\% &1.10\% &96.78\% &0.63\% &0.39\%\\
e &0.94\%  &0 &1.70\%  & 0.11\%  & 93.31\% & 0.94\%\\   
f &0.44\% &0.52\% & 0.31\%  &0.48\% &0.17\% & 98.08\%\\ \hline \hline
\end{tabular}
\label{dcnncon}
\end{table}
\section{Conclusions}

We introduced a simple and practical technique for effective automatic arm gesture recognition based on radar MD signature envelopes. No range or angle information was incorporated in the classifications. An energy-based thresholding algorithm was applied to separately extract the positive and negative frequency envelopes of the signal spectrogram. The extracted envelopes were concatenated and provided to different types of classifiers. We used the canonical angle to determine a prior whether the arm motions possess sufficient dissimilarities. The arm motion with the highest classification rate was selected as the "attention" motion to signaling the radar to begin and end sensing. It was shown that the NN classifier based on L1 distance achieves the highest accuracy and provided higher than 97 percent classification rate against various aspect angles and arm speeds. The experimental results also demonstrated that the proposed method outperforms handcrafted feature-based classification, all different forms of PCA-based classifications, and is comparable to the CNN method. It was also shown that the arm motion maximum instantaneous frequencies play a major role in the classification. The proposed automatic arm motion recognition method can be applied to control instrument and household appliances for smart home technology.


%





%

\bibliographystyle{IEEEtran.bst}
\bibliography{IEEEabrv,refs.bib}

%








\end{document}